\documentclass[aps,prl,showpacs,10pt,a4,nofootinbib,notitlepage,twocolumn,superscriptaddress]{revtex4-2}
\usepackage[utf8]{inputenc}
\usepackage[T1]{fontenc}
\usepackage{datetime}
\usepackage{dsfont}
\usepackage[colorlinks={},linkcolor=blue,citecolor=blue,urlcolor=blue,pdfauthor={ },pdftitle={ },pdfsubject={ },pdfkeywords={ }]{hyperref}
\usepackage{amssymb,amsmath,amsfonts,amsthm}
\usepackage{mathtools}
\usepackage{verbatim}
\usepackage{enumerate}
\usepackage{graphicx}
\graphicspath{{figs/}}
\usepackage{bbm}
\usepackage{booktabs}
\usepackage{ulem}
\usepackage{wasysym}

\usepackage[caption=false]{subfig}

\usepackage{color}

\usepackage{xcolor}

\definecolor{dred}{rgb}{.8,0.2,.2}
\definecolor{ddred}{rgb}{.8,0.5,.5}
\definecolor{dblue}{rgb}{.2,0.2,.8}
\definecolor{dgreen}{rgb}{.2,0.5,.2}

\newcommand{\ket}[1]{\ensuremath{|#1\rangle}}

\newcommand{\Tr}{\textrm{Tr}}

\begin{document}
	
\title{Thermal vapor quantum battery based on collective atomic spins}

\author{Jinyi Li}
\thanks{These authors contributed equally to this work.}
\affiliation{School of Physics and Optical Engineering, Zhejiang University of Technology, Hangzhou 310023, China}
\affiliation{Laboratory of Spin Magnetic Resonance, School of Physical Sciences, Anhui Province Key Laboratory of Scientific Instrument Development and Application, University of Science and Technology of China, Hefei 230026, China}

\author{Juncheng Zheng}
\thanks{These authors contributed equally to this work.}
\affiliation{School of Physics and Optical Engineering, Zhejiang University of Technology, Hangzhou 310023, China}

\author{Xue Yang}
\thanks{These authors contributed equally to this work.}
\affiliation{School of Information Science and Technology, Southwest Jiaotong University, Chengdu 610031, China}

\author{Kainan Hu}
\affiliation{School of Physics and Optical Engineering, Zhejiang University of Technology, Hangzhou 310023, China}

\author{Kanzheng Zhou}
\affiliation{School of Physics and Optical Engineering, Zhejiang University of Technology, Hangzhou 310023, China}

\author{Junkai Zhuang}
\affiliation{School of Physics and Optical Engineering, Zhejiang University of Technology, Hangzhou 310023, China}

\author{Hengyan Wang}
\email{hywang@zust.edu.cn}
\affiliation{Department of Physics, Zhejiang University of Science and Technology, Hangzhou 310023, China}

\author{Zhihao  Ma}
\email{mazhihao@sjtu.edu.cn}
\affiliation{School of Mathematical Sciences, MOE-LSC, Shanghai Jiao Tong University, Shanghai, 200240, China}

\author{Mingxing Luo}\email{mxluo@swjtu.edu.cn}
\affiliation{School of Information Science and Technology, Southwest Jiaotong University, Chengdu 610031, China}
\affiliation{Hefei National Laboratory, Hefei 230026, China}
 
\author{Wenqiang Zheng}
\email{wqzheng@zjut.edu.cn}
\affiliation{School of Physics and Optical Engineering, Zhejiang University of Technology, Hangzhou 310023, China}

	
\begin{abstract}

Quantum batteries harness non-classical resources, such as quantum coherence and entanglement, to surpass the performance limits of classical energy-storage devices. Here we realize a room-temperature quantum battery based on a collective atomic spin ensemble in a thermal alkali-metal vapor, containing approximately  $10^{12}$ $^{87}$Rb atoms with coherence times exceeding 110 ms. We operationally determine the battery capacity by directly measuring the extremal internal energies accessible under unitary evolution. This tomography-free protocol agrees closely with the conventional state-based definition and verifies the decomposition of capacity into coherent and incoherent contributions. We further show that quantum coherence can substantially enhance the storage capability independently of level populations, and experimentally establish quantitative relations linking battery capacity to von Neumann, Tsallis and linear entropies. By introducing a controlled dephasing channel with a magnetic-field gradient, we observe a monotonic reduction of capacity with coherence loss and track the corresponding evolution of the entropy–capacity relations. Our results identify thermal atomic spin ensembles as a scalable platform for quantum batteries and connect macroscopic quantum energy storage with operational quantum thermodynamics.
\end{abstract}

\maketitle

\noindent\textbf{Introduction}\\
The rapid advancement of quantum technology is creating an urgent demand for next-generation quantum devices \cite{campaioli2024colloquium,camposeo2025quantum}. In this context, quantum batteries have emerged as a novel energy-storage framework that utilizes quantum-mechanical principles \cite{allahverdyan2004maximal,alicki2013entanglement}. Unlike classical storage systems, quantum batteries aim to harness distinct resources such as quantum coherence, correlations, and entanglement to enhance performance in capacity, work extraction, and charging power \cite{Campaioli2017PhysRevLett,Shi2022PhysRevLett,perarnau2015extractable,andolina2019extractable,kamin2020entanglement,simon2025correlations}. Early theoretical explorations demonstrated the promise of quantum energy storage through models like Dicke battery \cite{ferraro2018high}. This inspired a variety of quantum battery designs, including Tavis–Cummings systems, spin chains, and strongly correlated SYK fermionic systems \cite{andolina2019extractable,le2018spin,PhysRevLett.125.236402,garcia2020fluctuations,ahmadi2024nonreciprocal,catalano2024frustrating,lu2025topological}. These schemes typically use engineered quantum interactions to achieve  scalable work extraction \cite{hovhannisyan2013entanglement,skrzypczyk2014work,richens2016work,Tirone2023PhysRevLett,salvia2023optimal,binder2015quantacell,song2024remote,andolina2025genuine} or  enhanced storage capacity \cite{julia2020bounds,francica2020quantum,Zhang2024PhysRevA}. Such performances establish quantum batteries as indispensable components for emerging quantum technologies \cite{Perarnau-Llobet2020NatRevPhys,horodecki2013fundamental,Alicki2019PhysRep,ganardi2025second} .

The core concepts of quantum batteries, such as collective charging, enhanced power, and work extraction, have been experimentally demonstrated in some controllable platforms, including superconducting circuits \cite{hu2022optimal}, trapped ions \cite{Zhang2025PhysRevLett}, semiconductor quantum dots \cite{maillette2023experimental}, single nitrogen-vacancy centers in diamond \cite{niu2024experimental}, and nuclear magnetic resonance (NMR) ensemble \cite{PhysRevA.106.042601}. While most demonstrations remain constrained to small numbers of addressable degrees of freedom, they often rely on highly engineered quantum states, short coherence times, precise quantum control, or cryogenic operation. Even in ensemble-based platforms such as NMR systems, the extractable work is typically limited by extremely weak spin polarization, yielding negligible usable energy per battery cycle and making many-body extensions 
particularly challenging, which is essential for further enhancing quantum battery performance 
~\cite{Campaioli2017PhysRevLett,kamin2020entanglement,alicki2013entanglement,shi2025quantum}.

We introduce a practical quantum battery platform based on a room-temperature collective atomic spin ensemble in a thermal alkali-metal vapor. Unlike existing quantum battery platforms \cite{hu2022optimal,Zhang2025PhysRevLett,maillette2023experimental}, atomic spin ensembles comprise a macroscopic number of quantum constituents and naturally support collectively controllable spin degrees of freedom at room temperature. The stored energy is directly from the collective spin polarization and scales extensively with the number of atoms. Optical pumping and radio-frequency (RF) control enable fast, reversible manipulation of the battery state, while non-destructive optical readout allows real-time monitoring of stored energy and capacity. Atomic spin ensembles have also been instrumental in the exploration of macroscopic quantum effects and 
generation of spin-squeezed and entangled states \cite{PhysRevLett.130.203602,bao2020spin,kong2020measurement,novikov2025hybrid,PhysRevLett.107.080503}. These well-established techniques make this platform an ideal testbed for investigating how different quantum resources enhance the performance of quantum batteries. 

We use a paraffin-coated antirelaxation cell to achieve near-unity spin polarization across approximately 10$^{12}$ atoms, with coherence times exceeding 110 ms—two orders of magnitude longer than those of typical room-temperature quantum systems. By applying a magnetic-field gradient to induce controllable decoherence, we achieve tunable control over the coherence of the spin ensemble. This allows us to experimentally demonstrate that quantum coherence enhances the storage capability independent of atomic populations. Furthermore, within this system, we experimentally observe explicit quantitative relationships among quantum coherence, entropy, and battery capacity \cite{allahverdyan2004maximal,yang2023battery}, thereby linking quantum information theory and thermodynamics in a macroscopic, experimentally accessible framework.

\begin{figure}[h]
\includegraphics[width=1\linewidth]{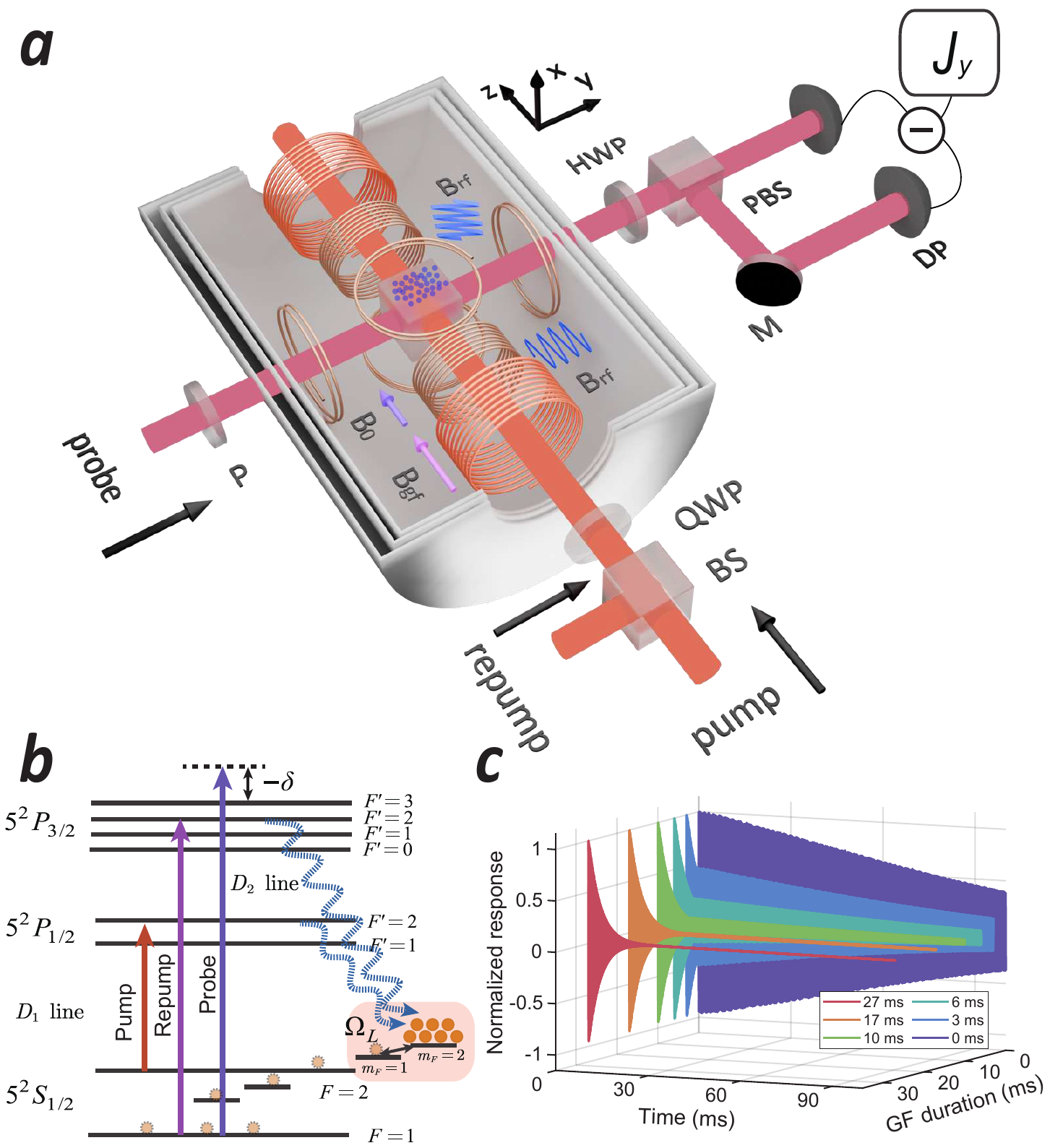}
\caption{Experimental platform for the thermal atomic-spin quantum battery. \textbf{a,} Schematic of the experimental apparatus. A paraffin-coated $^{87}$Rb vapour cell is driven by pump and repump beams for spin polarization, while a detuned probe beam enables non-destructive Faraday readout of the collective spin. HWP,  half-wave plate; QWP,  quarter-wave plate; PBS, polarizing beamsplitter; BS, beam splitter; M, mirror; P, polarizer; DP, differential photodiode. \textbf{b,} Relevant optical and Zeeman energy levels of $^{87}$Rb. The quantum battery is implemented in an effective two-level subspace formed by the Zeeman states $\ket{F=2, m_F=2}$ and $\ket{F=2, m_F=1}$. 
\textbf{c,} FID signals measured after application of gradient-field pulses of different durations.}
\label{Figure1}
\end{figure}

\bigskip
\noindent\textbf{Battery model with atomic spins}\\
We realize a quantum battery using a thermal ensemble of collective atomic spins composed of $N$ $^{87}$Rb atoms, which have been employed in quantum metrology or to explore macroscopic quantum phenomena \cite{bao2020spin}. We employ a coherently pumped, spin-polarized hot $^{87}$Rb atomic ensemble in a uniform biased magnetic field $\mathbf{B}$, where the Zeeman energy sublevels $\ket{F=2, m_F=2}$ and $\ket{F=2, m_F=1}$ form an effective two-level system. 

The state of the atomic spin ensemble is described by
\begin{equation}
\begin{split}
\begin{aligned}
 \hat{\varrho}=\frac{1}{2N}\sum_{n=1}^N{\left( \mathbb{I}+\sum_{k=x,y,z}^{}{\left< j_{k}^{n} \right>}\sigma _k \right)}=\frac{1}{2}\sum_{k=x,y,z}{\left( \mathbb{I}+\left< J_k \right> \sigma _k \right)}, 
\end{aligned}
\end{split}     
\end{equation}
where $\mathbb{I}$ denotes the identity matrix, $\sigma _k$ ($k=x, y, z$) are Pauli matrices, $j_{k}^{n}$ represents the $k$-component of the spin angular momentum of the $n$th atom, and $J_k = \frac{1}{N}\sum_{n=1}^N j_{k}^{n}$ is the normalized collective spin.
With the bias magnetic field $B_0$ applied along the optical pumping direction ($z$-axis), the ensemble is prepared in a near-fully polarized collective spin coherent state, with $\left< J_z \right> \approx N/2$ and $\left< J_x \right> = \left< J_y \right> = 0$.

The quantum system is governed by the Hamiltonian $H_S=\sum_n{\epsilon _n\left| n \right>}\left< n \right|$,  where $\epsilon _n=\gamma \hbar B_0/2$ is the energy eigenvalue associated with the eigenstate $\left| n \right>$, $\gamma$ is the gyromagnetic ratio of $^{87}$Rb atomic spins. The total internal energy of the system is given by $E=\Tr\left[ \hat{\varrho}H_S \right]$.
Cyclic unitary control of the collective spin is implemented by applying a transverse RF magnetic field, described by $\mathbf{ B}_{\rm rf}(t) = B_{\rm rf}\cos{(\omega_{\rm rf}t)}[{\cos\phi _{\rm rf}}\mathbf{e_x}+\sin\phi _{\rm rf}\mathbf{e_y}]$, where $B_{\rm rf}$ is the field amplitude, $\omega_{\rm rf}$ is the driving frequency, and $\phi_{\rm rf}$ determines the polarization orientation of the RF field in the $xy$-plane.

The energy-storage capability of a quantum battery is characterized by its ergotropy, which quantifies the maximum extractable work under cyclic unitary evolution, and its anti-ergotropy, which corresponds to the maximum injectable work  \cite{francica2020quantum,yang2023battery,julia2020bounds}. The total storage capacity is naturally defined as the sum of these two complementary quantities. From an operational perspective, the capacity is an intrinsic property that quantifies the full energetic range accessible to a quantum battery under quantum-adiabatic unitary transformations available in a closed quantum system. In atomic systems, the battery capacity is determined by the competition between the pump rate and the relaxation rate, as specified in the Eqs.\ref{eq8}-\ref{eqt:capacity with mean spin magnitude} in the Methods. The effects of various atomic decoherence processes and optical pumping on the battery capacity can also be found in the methods.

\begin{figure*}[t]
\includegraphics[width=17cm]{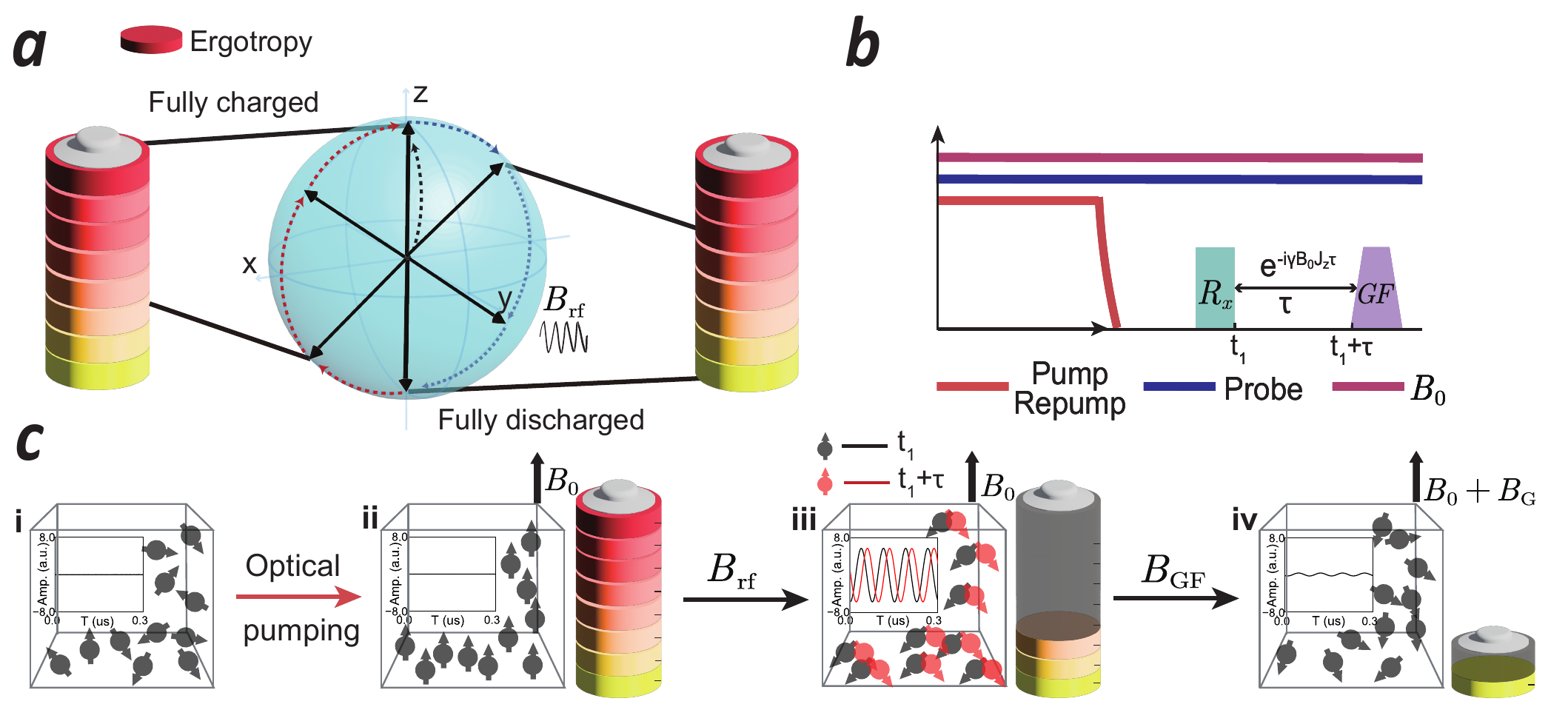}
\caption {Microscopic charge–discharge cycle and experimental control sequence. \textbf{a,} 
Schematic of a microscopic charge–discharge cycle on the Bloch sphere. The north and south poles correspond to the fully charged and fully discharged battery states, respectively. Blue dashed trajectories indicate discharge, corresponding to ergotropy reduction and work extraction, whereas red dashed trajectories indicate recharge and work injection. The black dashed path denotes battery-state preparation. 
\textbf{b,} After adiabatic switch-off of the optical pumping fields, the initial battery state is prepared by coherent spin rotations. Subsequent unitary control operations, together with optional gradient-field pulses, implement charging, discharging and dephasing, followed by optical readout. Rotations about the 
$x$- and $y$-axes are driven by transverse RF pulses, whereas rotations about the 
$z$-axis are realized by free precession under the bias magnetic field for a controlled duration.
\textbf{c,} Schematic evolution of the atomic state, battery capacity and ergotropy during the experimental sequence. The arrows inside the atomic spheres denote the direction of collective spin polarization, while the size and coloured fraction of the battery icons represent the total capacity and ergotropy, respectively. Insets show representative measured time-domain readout signals after the applied control sequence.
}
\label{Figure22}
\end{figure*} 

\bigskip
\noindent\textbf{Experiments}\\
\noindent\textit{Experimental setup}\\
The experimental setup is shown in Fig. \ref{Figure1}\textbf{a}. A  paraffin-coated $^{87}$Rb vapor cell with dimensions of $2\times2\times2$ cm$^3$ serves as the quantum battery medium. The cell temperature is actively stabilized at 320.15 K to increase the atomic density while remaining below the melting temperature of the paraffin coating. The paraffin coating effectively suppresses wall-collision-induced decoherence, leading to extended longitudinal ($T_1$) and transverse ($T_2$) relaxation times ($T_1 = 220.8$ ms and $T_2 = 113.4$ ms, Fig. \ref{Figure1}\textbf{c}). These coherence times ensure
long lifetime of the quantum energy states and
are much longer than the maximum pulse operation duration ($\le 3.6$ ms), thereby enabling nearly dissipation-free quantum control.

The atomic spin ensemble is controlled using three key subsystems.
(i) Optical system. A pump laser locked to the Rb D1 transition $5S_{1/2}F=2 \to  5P_{1/2}F^{'}=2$ and a repump laser stabilized to the Rb D2 transition $5S_{1/2}F=1 \to 5P_{3/2} F^{'}=2$, both with $\sigma^+$ circular polarization, drive the ensemble into a near-unity spin-polarized state (yielding a polarization of $\approx 0.99(4)$) along the $z$-axis. The pump and repump beams are switched using acousto-optic modulators. A linearly polarized probe laser, blue-detuned by 2.3 GHz from the $5S_{1/2}F=2 \to 5P_{3/2}F'=3$ transition, enables quantum non-destructive measurement of the transverse spin component $J_y$ via Faraday rotation. 
For atomic state tomography, free-induction-decay (FID) signals are recorded using the probe detection. The coherence terms of the density matrix are extracted by fitting the Fourier-transformed FID spectra. The diagonal (population) terms are obtained by applying a $\pi/2$ pulse to rotate the $J_z$ component into the transverse plane, where it is subsequently read out by the probe light.
The corresponding optical transition energy-level diagram is shown in Fig. \ref{Figure1}\textbf{b}. (ii) Static magnetic-field system. The vapor cell is housed inside a five-layer cylindrical magnetic shield, providing passive isolation of the atomic spins from complex ambient magnetic field disturbances. A set of coils inside the shield generates a uniform static magnetic field, which defines the energy splitting between the two extremal atomic spin states. In addition, a pair of anti-Helmholtz coils produces a magnetic field gradient along the $z$-axis, introducing a controlled source of dephasing for the atomic spin ensemble. As shown in Fig.~\ref{Figure1}\textbf{e}, the coherence magnitude decays monotonically with increasing duration of the applied gradient field (GF), whereas the population distribution remains unchanged. (iii) Coherent spin-control system. Unitary rotations about the $x$ and $y$ axes are implemented by applying transverse RF magnetic-field pulses that coherently manipulate the collective atomic spin. These pulses are generated by two mutually orthogonal pairs of coils mounted around the cell. Arbitrary waveform generators produce RF pulses with fixed duration and variable amplitude, enabling rotations by arbitrary angles. Rotations about the 
$z$-axis are realized through free precession of the spin ensemble under the system Hamiltonian.

\begin{figure*}[t]
\includegraphics[width=17cm]{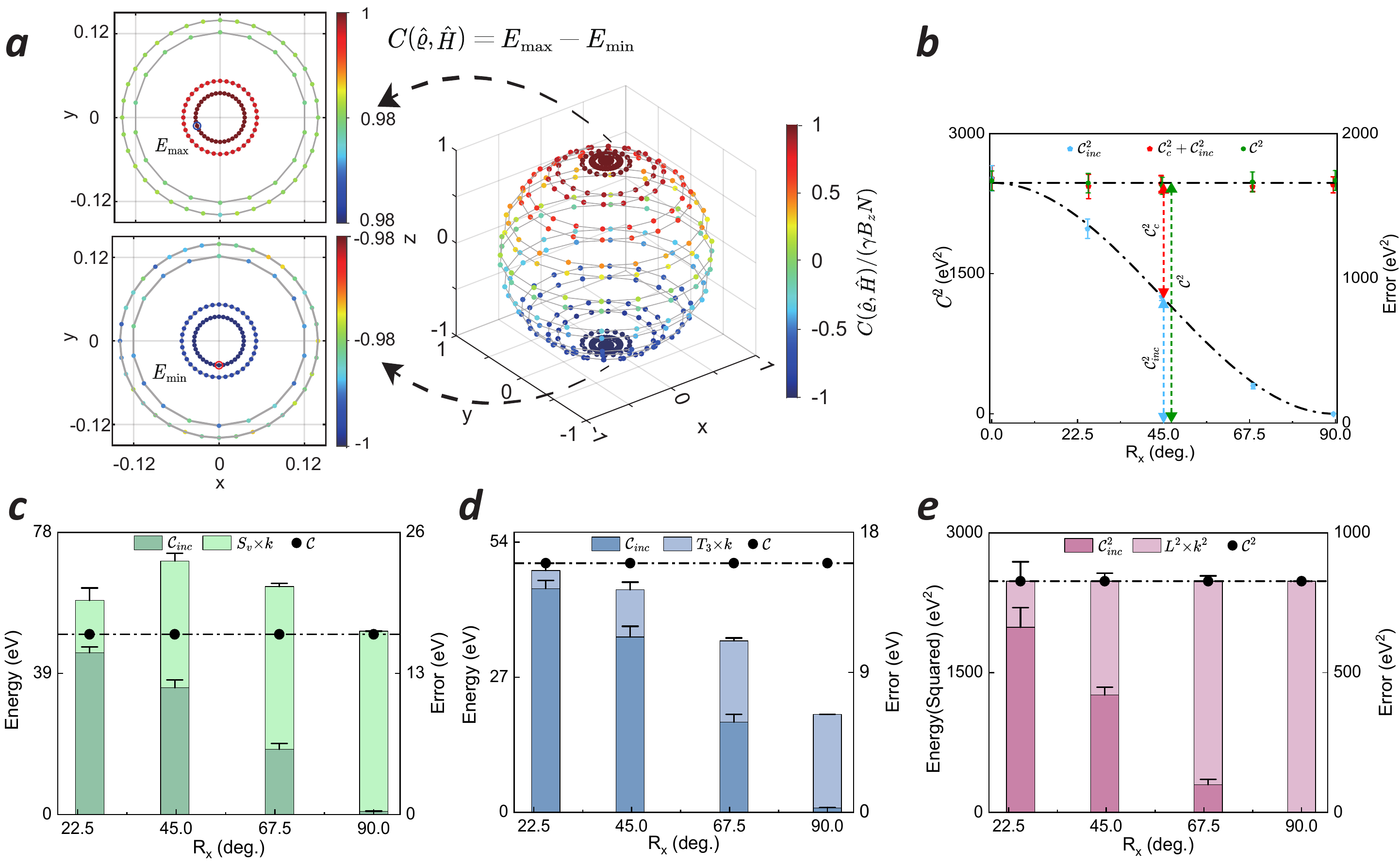}
\caption {
Experimental determination of quantum battery capacity and its entropic constraints. \textbf{a,} Capacity measurement by hierarchical traversal of the unitary control space. Starting from a state prepared by $R_z(200^\circ)R_x(33^\circ)$ after optical pumping, the accessible extremal energies are identified by sweeping unitary rotations. Insets show two-dimensional projections of the search space around the maximum and minimum.
\textbf{b,} Comparison of battery capacities obtained by different methods. The total capacity measured by hierarchical traversal is compared with the value inferred from state tomography, together with its coherent and incoherent contributions. The data confirm both the consistency between operational and state-based definitions of capacity and the Pythagorean decomposition $\mathcal{C}^2 = \mathcal{C}_{\mathrm{c}}^2 + \mathcal{C}_{\mathrm{inc}}^2$.
From the reconstructed density matrix, the capacity is obtained from its eigenvalues $\lambda_\pm$ as $\mathcal{C}=(\lambda_+-\lambda_-)N\gamma\hbar B_0/2$. Dotted lines indicate the corresponding theoretical predictions.
\textbf{c-e,} Constraint relations between battery capacity and the von Neumann entropy (\textbf{c}), Tsallis entropy (\textbf{d}), and linear entropy (\textbf{e}) of the system. Results are shown for both the coherent battery state and the corresponding incoherent state obtained after complete dephasing. Here $k=\hbar\gamma B_0 N$.}
\label{Figure33}
\end{figure*}

\bigskip
\noindent\textit{Atomic battery evolves under quantum controls}\\
Figure  \ref{Figure22}\textbf{a} shows one microscopic charge-discharge cycle. Unlike conventional approaches that rely on projective two‑point measurements or quasi‑static assumptions \cite{an2015experimental,batalhao2014experimental}, our method directly determines the energies of fully discharged and fully charged states via coherent unitary rotations. This enables a characterization of the full energy-storage capacity without invoking external work reservoirs or inducing state collapse. The key quantities-ergotropy (maximal extractable work) and antiergotropy (maximal injectable work) \cite{allahverdyan2004maximal,yang2023battery}-are controlled by transverse RF pulses that unitarily rotate the atomic spins while preserving the Hamiltonian spectrum. In the polarization stage, optical pumping prepares the battery in the fully charged state $|\!\!\uparrow\rangle$ aligned with the $+z$ direction. Along the blue trajectory shown in the figure, the population imbalance between the two energy eigenstates gradually vanishes and subsequently reverses. At the special point corresponding to a $\pi/2$ rotation, the extractable work originates entirely from quantum coherence. Upon completion of a $\pi$ rotation, the spin populations are fully inverted, marking the transition from the fully charged to the fully discharged state. Throughout this evolution, the unitary rotations reduce the ergotropy and extract work. Subsequent pulses restore the ergotropy, reinjecting work and completing the cycle, as indicated by the red trajectory. Because the capacity is invariant under unitary evolution \cite{yang2023battery}, the maximum and minimum attainable internal energies directly reveal the system’s work capacity, enabling inherently cyclic, far‑from‑equilibrium operation without quasi‑static constraints. 

In the experiment, the state of the quantum battery is jointly controlled by optical and magnetic fields. A representative timing sequence and the corresponding spin dynamics, together with the associated evolution of the battery state, are shown in Fig.~\ref{Figure22}\textbf{b} and \textbf{c}, respectively. The bias magnetic field and the probe beam are turned on throughout the entire sequence. The optical detection signal recorded after each operation is displayed in the inset. (i) For an unpolarized ensemble ($\left< J \right> =0$), both the capacity and ergotropy are zero. (ii) After a sufficient optical pumping period (full polarization), the capacity and ergotropy reach $49.8(7)$ eV, while the transverse signal exhibits only fluctuations around zero. (iii) The subsequent quantum-control stage implements a unitary rotation via a transverse RF pulse applied along the $x$ axis, generating a transverse spin component. Accordingly, the quantum battery evolves from the fully charged state to a chargeable state, leading to a reduction of ergotropy while the capacity remains invariant. A following free evolution, corresponding to spin precession about the $z$ axis, modifies the phase of the transverse component while conserving both capacity and ergotropy. (iv) Finally, application of a magnetic-field gradient induces controlled dephasing of the transverse component, leaving $J_z$ intact. This accelerates decoherence of the spin polarization and increases the mixedness of the collective spin state.

\bigskip
\noindent\textit{Verification of coherence-enhanced quantum battery}

Beyond the classical contribution arising from population imbalance, coherence between energy eigenstates can enlarge the energetically accessible range under unitary operations, thereby enhancing the battery capacity(See methods for details). This coherence-enabled contribution reflects intrinsically quantum features of the state and distinguishes quantum batteries from their classical counterparts. Moreover, from a thermodynamic perspective, the capacity is fundamentally constrained by the entropic structure of the state. Because entropy quantifies mixedness and bounds the degree of population ordering achievable through unitary transformations, it imposes intrinsic limits on both maximal extractable and injectable work. Establishing the quantitative relationship between capacity and entropy therefore provides insight into the thermodynamic resources underlying quantum energy storage.

To experimentally investigate the relationship between capacity and quantum coherence, as well as its connection to entropy, we implement an operational protocol that systematically explores the unitary control space and directly determines the extremal internal energies. Unlike approaches based on full state reconstruction, where capacity is inferred from the density matrix via numerical post-processing, the unitary-manifold sampling protocol realizes the optimization inherent in the definition of capacity at the experimental level. This relies only on coherent control and energy readout, and scales favorably to many-body systems. 
The pulse sequence used in the experiment is shown in Fig.\ref{pulsesequence1} in  methods.

As shown in Fig.~\ref{Figure33}\textbf{a}, a hierarchical scanning strategy is employed to streamline the experimental determination of the quantum battery capacity. A combination of rotations about the $x$, $y$, and $z$ axes is implemented to search for the extremal energies. A coarse scan with a $20^\circ$ step size is first performed to identify the approximate extremal region. A subsequent fine scan, using smaller increments of $5^\circ - 10^\circ$, is then carried out in the vicinity of the initially located extremum within a reduced search range, enabling a more precise determination of the extremal point. In the experiment, a set of battery states is prepared by applying rotations of varying angles about the $x$-axis to the spin coherent state initially polarized along the $z$-axis. To assess the effectiveness of the hierarchical method, we set the state-preparation rotation angle as $R_z(200^\circ)R_x(33^\circ)$, for which the scanning strategy is expected to produce nearly the largest deviation in identifying the energy extrema. A deviation of 0.35 eV from the theoretical value ($49.72$ eV) is observed for this case, corresponding to a relative deviation of $0.70\%$. In comparison, when we set $R_z(200^\circ)R_x(40^\circ)$ for state-preparation,  under which the extrema can be accurately located, the deviation reduces to $0.53\%$, which mainly reflects intrinsic experimental imperfections. In the vicinity of the extrema, the internal energy of the atomic ensemble is only weakly sensitive to the distance between the quantum state and the extremal state. This flatness provides an intrinsic reason for the small relative deviation of the scanning method.

The measured capacities are compared with values calculated from the eigenvalues of the battery’s density matrix reconstructed via state tomography \cite{yang2023battery}. The results shown in Fig.~\ref{Figure33}\textbf{b} demonstrate the consistency between the operational definition of quantum battery capacity and its state-based definition. The two methods yield consistent results, with deviations below 2\%. The Pythagorean relation of quantum capacity \eqref{Pythagorean} is also experimentally verified. The  coherent ($\mathcal{C}_{\mathrm{c}}$) and incoherent ($\mathcal{C}_{\mathrm{inc}}$) are obtained from the reconstructed density matrix. For an initial rotation angle of $25^\circ$, we measure $\mathcal{C} = 49.37$ eV, $\mathcal{C}_c = 21.36$ eV, and $\mathcal{C}_{\text{inc}} = 44.58$ eV. The agreement between experiment and the Pythagorean relation demonstrates that both coherence and level populations contribute to energy storage, and that the quantum capacity combines quadratically rather than linearly, which is a distinctive feature of coherent energy storage. Notably, when the total capacity is entirely contributed by coherence, the battery operates in a purely coherence-driven regime that is inaccessible to classical systems. These results confirm that coherence alone can support substantial energy storage.

Fig.~\ref{Figure33}\textbf{c–e} show the energy storage capacity as a function of the quantum entropy of the system. These results demonstrate that the energy stored in both the total state and the incoherent state (the state $\hat{\varrho}_{\rm{inc}}$ obtained after complete decoherence) is microscopically constrained by the corresponding information-theoretic entropies (See methods for more information). Since all prepared states are pure coherent states, their von Neumann, Tsallis, and linear entropies vanish, leading to equality in Eqs.~(\ref{von_entropy},\ref{Tsallis_entropy},\ref{linear_entropy}) for the total states. In contrast, for the decohered state $\hat{\varrho}_{\rm{inc}}$, the same relations reveal a trade-off: lower entropy allows higher capacity, as described by Eq.~\eqref{von_entropy}. Fig.~\ref{Figure33}\textbf{d} shows a complementary relation involving the Tsallis entropy (Eq.~\eqref{Tsallis_entropy}), characteristic of non-extensive thermodynamics. Figure~\ref{Figure33}\textbf{e} connects the incoherent capacity to the system purity (linear entropy) \cite{horodecki2009quantum} through Eq.~\eqref{linear_entropy}, quantifying how purity constrains energy storage. Experimental errors remain below 2.3\%, confirming the theoretical predictions for the incoherent states.

\begin{figure}[h]
\includegraphics[width=1\linewidth]{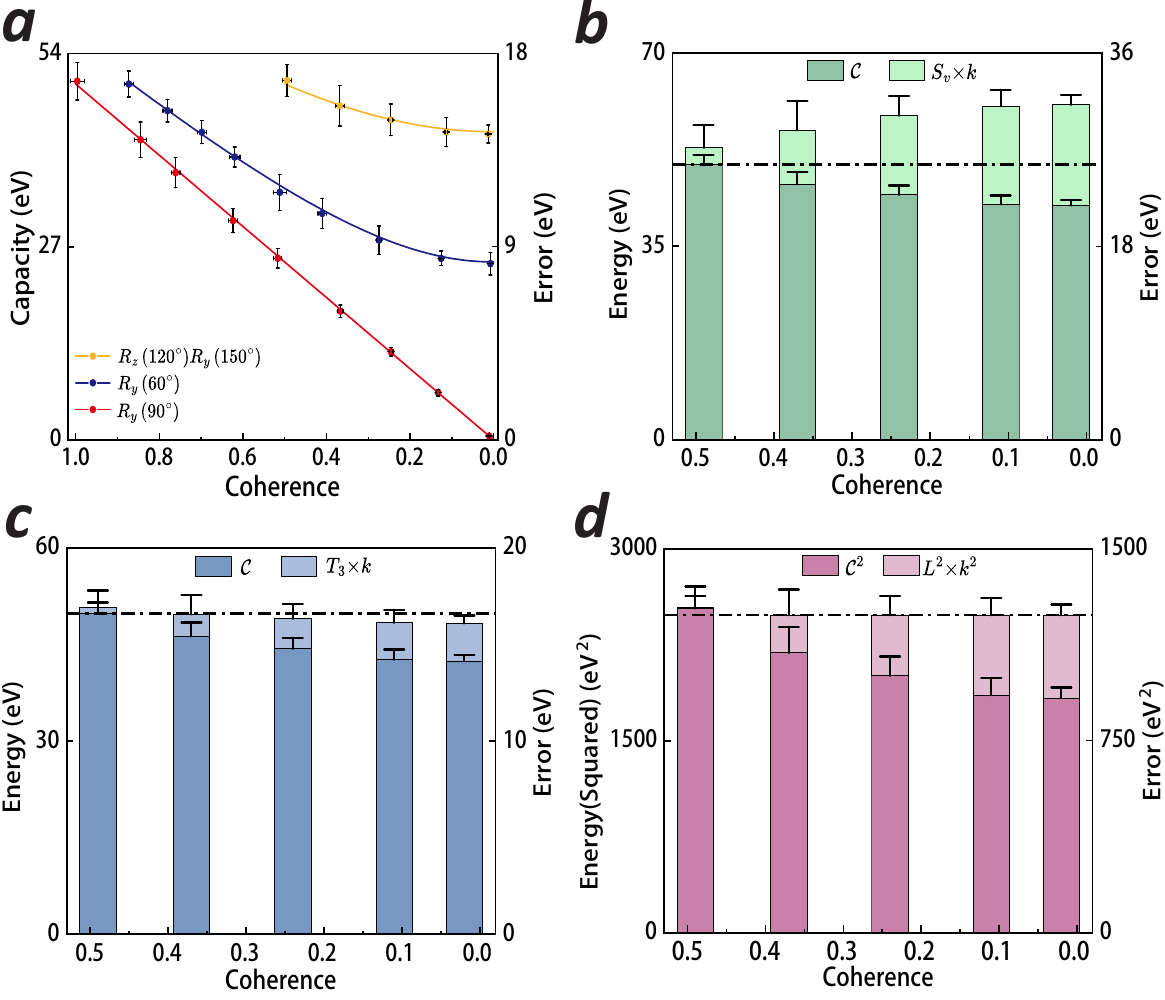}
\caption {Capacity degradation under controlled dephasing. \textbf{a,} Capacity as a function of coherence for three representative initial states. \textbf{b–d,} Evolution of the entropy–capacity relations during dephasing for the initial state for the initial state prepared by the rotations $R_z(120^\circ)R_y(150^\circ)$, Shown are the relations involving the von Neumann entropy (\textbf{b}), Tsallis entropy (\textbf{c}), and linear entropy (\textbf{d}) 
The data demonstrate that the capacity–coherence–entropy relations remain valid during decoherence and reveal how coherence modifies the approach to the corresponding entropic bounds. Points are averages over 15 independent runs and error bars denote the standard errors.}
\label{Figure44}
\end{figure}

\begin{figure}[h]
\includegraphics[width=1\linewidth]{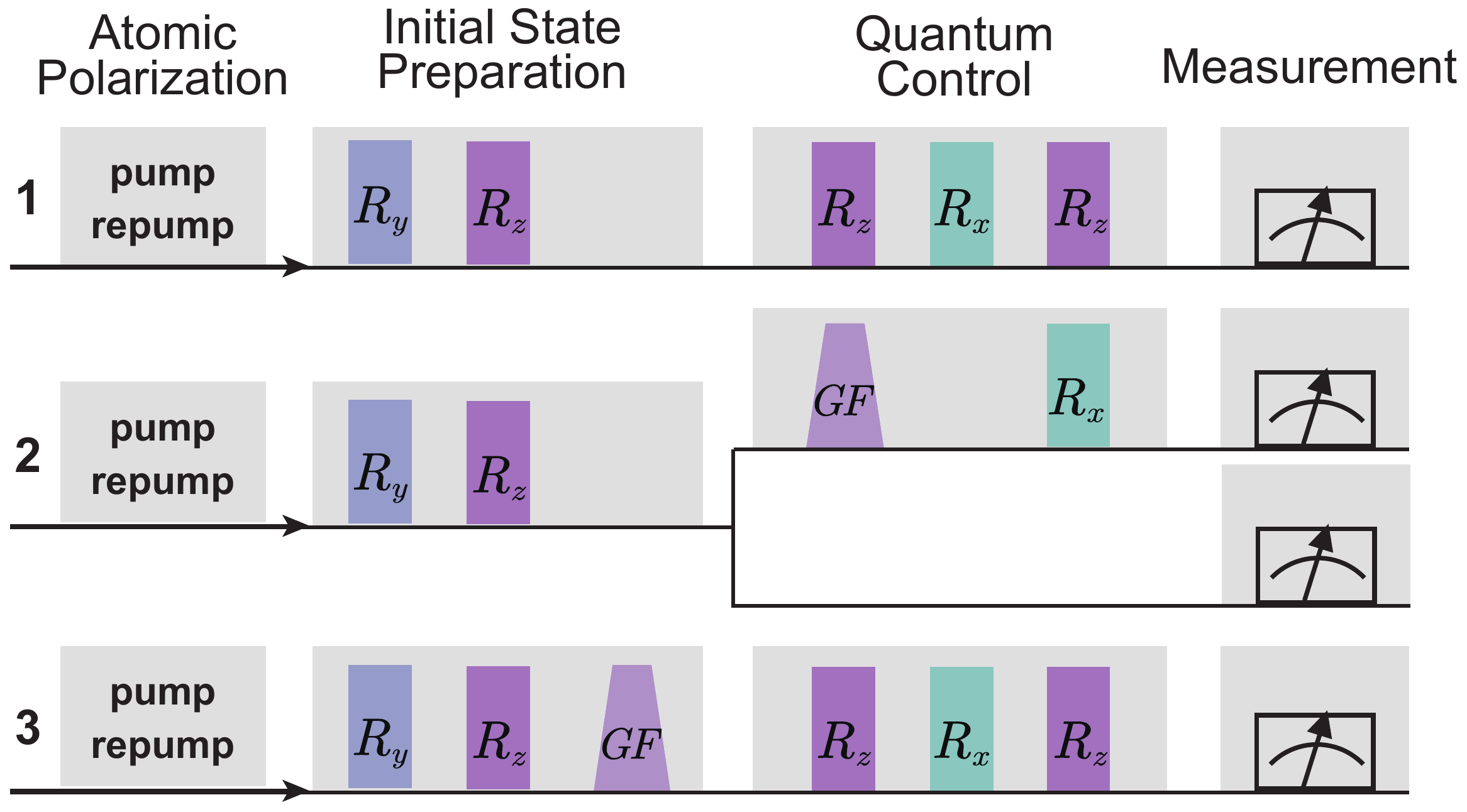}
\caption{Experimental protocols for capacity measurements. \textbf{1}, Hierarchical traversal for a coherent quantum battery. \textbf{2}, Tomographic reconstruction of capacity. \textbf{3}, Hierarchical traversal for an incoherent (classical) battery. Coloured blocks denote RF control pulses about different axes and gradient-field pulses.}
\label{pulsesequence1}
\end{figure}

\bigskip
\noindent\textit{Quantum capacity under engineered dephasing}\\
Quantum coherence is generally fragile under decoherence in quantum systems, and its degradation leads to a reduction in quantum battery capacity. To investigate this effect, different battery states, prepared by applying different rotations after the optical pumping stage, are subjected to a dephasing channel engineered by a spatial magnetic-field gradient. For each initial state, the coherence terms at different times during the dephasing process are reconstructed via partial state tomography, while the corresponding battery capacity is determined using the hierarchical scanning method. As shown in Fig.~\ref{Figure44}\textbf{a}, the measured coherence terms and the corresponding capacities demonstrate that the quantum capacity decreases monotonically as the coherence is reduced. For the prepared maximally coherent state, the coherence decays from 0.996 to 0.011 under the dephasing channel, accompanied by a decrease in capacity from $5.00$ eV to $0.54$ eV. Different initial rotations give rise to distinct decay profiles: only the state prepared by the rotation $R_y(90^\circ)$ exhibits an approximately linear dependence, whereas the others display nonlinear behavior. This indicates a state-dependent efficiency in the utilization of coherence.

Figures~\ref{Figure44}\textbf{b–d} show the influence of coherence on the entropic behavior and demonstrate that the capacity–coherence–entropy relations remain valid during the decoherence of a state prepared by the rotations $R_z(120^\circ)R_y(150^\circ)$. Compared with the results for incoherent states in Figs.~\ref{Figure33}\textbf{c–e}, these data reveal the open-system dynamics of a coherent battery coupled to the environment. In the presence of coherence, the linear combinations of the battery capacity and the entropy, weighted by the coefficient appearing in the inequalities, approach the bounds on the right-hand side more closely. For the von Neumann entropy relation, the deviations from the bound are reduced from  $1.8\%$-$40\%$ to $0.6\%$-$2.2\%$, while for the Tsallis entropy relation they decrease from  $3\%$-$60.7\%$ to $0.7\%$-$2\%$. This indicates that the presence of coherence suppresses entropy production and enhances energy-conversion efficiency.
The data further show that the degradation of coherence is accompanied by a reduction in capacity. While the bound involving the von Neumann entropy becomes less tight as coherence decreases (Fig.~\ref{Figure44}\textbf{b}), the bounds based on the Tsallis entropy (Fig.~\ref{Figure44}\textbf{c}) and the linear entropy (Fig.~\ref{Figure44}\textbf{d}) remain tight within experimental uncertainty ($<3.9\%$). 

\bigskip
\noindent\textbf{Conclusions and outlook}
\\
We have realized a quantum battery based on a room-temperature collective atomic spin ensemble in a thermal alkali-metal vapor, and established an operational framework for quantifying its energy-storage capability. By combining coherent optical pumping, RF spin control and non-destructive optical readout, we directly determined the extremal internal energies accessible under unitary evolution, thereby measuring the battery capacity without full state reconstruction. The measured capacities agree closely with the values inferred from state tomography, confirming the consistency between the operational and state-based definitions of quantum battery capacity. We further verified the Pythagorean decomposition of the capacity into coherent and incoherent contributions, demonstrating that quantum coherence and level populations contribute in fundamentally different ways to energy storage. Our experiments also reveal how quantum coherence shapes the thermodynamic performance of the battery. We show that coherence alone can support substantial capacity, and we experimentally establish quantitative relations linking capacity to von Neumann, Tsallis and linear entropies. The observed trade-offs between capacity and entropy further suggest thermodynamic uncertainty relations, extending the connection between quantum-state structure and work extraction beyond classical energy-storage paradigms.
By introducing a controlled dephasing channel, we directly observe the monotonic reduction of capacity with coherence loss and track the corresponding evolution of the entropy–capacity relations. These results identify thermal atomic spin ensembles as a versatile platform for studying the interplay between coherence, mixedness and energy in a macroscopic quantum system.

Looking ahead, this platform offers several natural extensions. the rich hyperfine and Zeeman structure of alkali atoms offers a route toward multilevel and higher-dimensional quantum batteries, while more elaborate control sequences may enable access to non-classical states such as spin-squeezed or entangled collective states. It should also be possible to explore non-equilibrium thermodynamic protocols beyond the near-unitary regime, including controlled dissipation, feedback-assisted charging. More broadly, by combining room-temperature operation, large atom number, long coherence time and operationally meaningful measurements, this system opens a path toward experimentally scalable quantum batteries and toward a deeper understanding of macroscopic quantum energy storage.

\bigskip
 \noindent\textbf{Methods}\\
 \noindent\textit{Battery capacity}\\
For a closed quantum system undergoing unitary dynamics, the ergotropy quantifies the maximum work extractable from a state, while the anti-ergotropy gives the maximum work required to create that state from its passive counterpart. Consistent with the first law of quantum thermodynamics \cite{strasberg2021first}, ergotropy (anti-ergotropy) corresponds to the largest decrease (increase) in the internal energy achievable by unitary operations on the state.

The total battery capacity, defined as the sum of ergotropy and anti-ergotropy \cite{yang2023battery}, measures the complete energetic controllability of the system. Equivalently, it equals the difference between the maximum and minimum internal energies attainable  under cyclic unitary transformations, thus capturing the full range of reversible energy variations. For an ensemble of $N$ non-interacting spin-1/2 particles (or two-level atoms) in a uniform magnetic field $B_0$ along the $z$-axis, the capacity reads
\begin{eqnarray}
 \mathcal{C }(\hat{\varrho } ,\hat{H_0} ) 
 &=& \max_{U\in \mathbb{SU}(2)}{\rm Tr}(U\hat{\varrho} U^\dagger H)-\min_{U\in \mathbb{SU}(2)}{\rm Tr}(U\hat{\varrho} U^\dagger H)
 \nonumber \\
  &=& 2\hbar\gamma B_0\sqrt{\bigl(\sum_{n=1}^N \langle{j_x^n}\rangle\bigr)^2+\bigl(\sum_{n=1}^N \langle{j_y^n}\rangle\bigr)^2+\bigl(\sum_{n=1}^N \langle{j_z^n}\rangle\bigr)^2}
   \nonumber.
 \label{capacity}
\end{eqnarray}

Solving the master equation (\ref{eq2}) for the effective atomic two-level system yields the battery capacity
\begin{equation}
\mathcal{C}(\hat{\varrho},\hat{H}_0)
=
\hbar \gamma B_0 N
\frac{R_{\mathrm{op}}}{R_{\mathrm{op}}+R_{\mathrm{rel}}}
\left[1-e^{-(R_{\mathrm{op}}/2+R_{\mathrm{rel}})t}\right],
\label{eq8}
\end{equation}
where \(R_{\mathrm{rel}}\) denotes the total relaxation rate. In our experiment, the optical-pumping duration is much longer than the characteristic relaxation times, so the system reaches a steady state and the transient exponential term can be neglected. The steady-state capacity then reduces to
\begin{equation}
\mathcal{C}_{\mathrm{ss}}=\hbar \gamma B_0 N \tanh\!\left(\frac{\beta}{2}\right),
\label{eqt:spin temperature}
\end{equation}
where the spin temperature $\beta$ characterizes the degree of atomic polarization. It is related to the measured polarization $P$ through
$\beta=\ln\left(\frac{1+P}{1-P}\right)$.
In the high-polarization limit, the capacity further simplifies to
\begin{equation}
\mathcal{C}_{\mathrm{ss}} \approx \hbar \gamma B_0 N\,|\bar{j}|,
\label{eqt:capacity with mean spin magnitude}
\end{equation}
where \(|\bar{j}|\) denotes the mean spin magnitude per atom. This expression highlights that the accessible energy-storage capacity is jointly determined by the degree of collective spin alignment and the size of the atomic ensemble.

\bigskip
\noindent\textit{Coherence-resolved battery capacity}\\
A more refined energetic characterization is obtained by decomposing the battery capacity into coherent and incoherent contributions with respect to the energy eigenbasis, following the resource-theoretic framework of quantum coherence \cite{baumgratz2014}. Within this picture, the total battery capacity naturally separates into two distinct parts: a coherent capacity $\mathcal{C}_c$ and an incoherent capacity $\mathcal{C}_{\rm inc}$.

For a general spin state $\hat{\varrho}$, we decompose the density matrix into a coherent part $\hat{\varrho}_c$ and an incoherent part $\hat{\varrho}_{\text{inc}}$ with respect to the energy eigenbasis of the local Hamiltonian $\hat{H}_0 = \hbar\gamma B_0 \sum_n \hat{j}_z^n$. The coherent part is written as
\begin{eqnarray}
\hat{\varrho}_c = \frac{1}{2N}\sum_{n=1}^N (\mathbb{I} + \langle j_x^n \rangle \sigma_x +\langle  j_y^n \rangle\sigma_y),
\end{eqnarray}
which leads to the coherent capacity
\begin{eqnarray}
\mathcal{C}^2_c(\hat{\varrho}_c, \hat{H}_0) = (2\hbar\gamma B_0 )^2\bigl[ \bigl(\sum_{n=1}^N \langle j_x^n\rangle \bigr)^2 + \bigl(\sum_{n=1}^N \langle j_y^n\rangle \bigr)^2\bigr].
\label{eq:choerent capacity}
\end{eqnarray}

The incoherent part contains only populations in the energy eigenbasis, 
\begin{eqnarray}
\hat{\varrho}_{\text{inc}} = \frac{1}{2N}\sum_{n=1}^N (\mathbb{I} + \langle j_z^n\rangle \sigma_z^n),
\end{eqnarray}
with the associated incoherent capacity
\begin{eqnarray}
\mathcal{C}^2_{\text{inc}}(\hat{\varrho}_{\text{inc}}, \hat{H}_0) = \bigl(2\hbar\gamma B_0\sum_{n=1}^N \langle j_z^n\rangle \bigr)^2.
\label{eq:inchoerent capacity}
\end{eqnarray}
 This represents the classical-like contribution to the battery capacity that remains even in the absence of coherence.


Remarkably, these two contributions satisfy a Pythagorean relation,
\begin{eqnarray}
\mathcal{C}^2(\hat{\varrho}, \hat{H}_0) = \mathcal{C}_c^2(\hat{\varrho}_c, \hat{H}_0) + \mathcal{C}_{\text{inc}}^2(\hat{\varrho}_{\text{inc}}, \hat{H}_0),
\label{Pythagorean}
\end{eqnarray}
reflecting the orthogonality between the transverse coherent components and the longitudinal population component on the Bloch sphere. This decomposition makes explicit how quantum coherence enlarges the accessible battery capacity, providing an operational signature of quantum advantage over any classical counterpart lacking coherent resources.

\bigskip
\noindent\textit{Entropy–capacity relations}\\
The storage capacity of a quantum battery is fundamentally constrained by information-theoretic quantities that characterize the disorder, purity, and information content of the state \cite{yang2023battery}. For $N$ non‑interacting spin‑$1/2$ particles, one can derive complementary relations between energetic resources and several entropy measures.

The von Neumann entropy, $S_v(\hat{\varrho}) = -\mathrm{Tr}(\hat{\varrho} \log_2 \hat{\varrho}) $, defined using the base-2 logarithm, quantifies the quantum information content of the state. It satisfies the following inequality with the battery capacity:
\begin{eqnarray}
\mathcal{C}(\hat{\varrho}, \hat{H}_0) + S_v(\hat{\varrho})\, \hbar\gamma B_0 N \ge \hbar\gamma B_0 N .
\label{von_entropy}
\end{eqnarray}
This relation expresses a trade-off between energy-storage capacity and information-theoretic uncertainty: increasing von Neumann entropy, corresponding to greater mixedness, constrains how small the capacity can become. The bound is saturated in two limiting cases, namely pure states with zero entropy and the completely mixed state with maximal entropy for a fixed Hilbert-space dimension.

Beyond the Shannon‑type entropy, the Tsallis entropy of order $p\ge 2$ provides a one-parameter family of non-extensive entropy measures \cite{tsallis1988possible},
\begin{eqnarray}
T_p(\hat{\varrho}) = \frac{1 - \lambda_{-}^p - \lambda_{+}^p}{p-1},
\end{eqnarray}
where $\lambda_{\pm} = (1 \pm S)/2 $ are the eigenvalues of the single-spin density matrix, and $S$ denotes the Bloch-vector length. In this case, the Tsallis entropy satisfies the upper-bound relation
\begin{eqnarray}
\mathcal{C}(\hat{\varrho}, \hat{H}_0) + T_p(\hat{\varrho})\, \hbar\gamma B_0 N \le \hbar\gamma B_0 N,
\label{Tsallis_entropy}
\end{eqnarray}
which becomes tighter as $p$ increases, reflecting the different sensitivity of generalized entropies to state mixedness.

An exact relation can also be established between the battery capacity and the system purity, quantified by the linear entropy $ L^2(\hat{\varrho}) \equiv T_2(\hat{\varrho}) $ \cite{horodecki2009quantum}. In this case, the following identity holds:
\begin{eqnarray}
\mathcal{C}^2(\hat{\varrho}, \hat{H}_0) + 2\, L^2(\hat{\varrho})\, (\hbar\gamma B_0 N)^2 = (\hbar\gamma B_0 N)^2 .
\label{linear_entropy}
\end{eqnarray}
This Pythagorean-type relation shows that the squared battery capacity and the appropriately scaled mixedness sum to a constant set by the external magnetic field and the system size. Taken together, these entropy–capacity relations provide a unified information-theoretic perspective on the fundamental limits of quantum batteries.

\bigskip
\noindent\textit{Dissipative dynamics of the atomic-spin battery}\\
The microscopic Hamiltonian of the $^{87}$Rb ground-state manifold is 
\begin{equation}
H_g=A_g\mathbf{I} \cdot \mathbf{S}+g_{s}\mu_{B}\mathbf{S}\cdot\mathbf{B} -\frac{\mu_{I}}{I}\mathbf{I}\cdot\mathbf{B},
\end{equation}
which determines the hyperfine and Zeeman level structure. Here $A_g=2\hbar\Delta_{hf}/(2I+1)$ is the magnetic-dipole hyperfine coupling strength between the nuclear spin $\mathbf{I}$ and the electron spin $\mathbf{S}$, $\Delta_{hf}$ is the hyperfine splitting frequency
$I=3/2$ for $^{87}$Rb, $g_{s}$ is the Landé factor, and $\mu_{B}$ and $\mu_{I}$ denote the Bohr magneton and nuclear magnetic moment, respectively. 


During free evolution, the atomic quantum battery is subject to several decoherence mechanisms, analogous to capacity degradation in classical batteries. Taking into account atomic diffusion within the vapor cell, the dynamics are described by the master equation
\begin{equation}
\begin{split}
\begin{aligned}
\frac{\mathrm{d}}{\mathrm{d} t}\hat \rho=& \frac{1}{i\hbar}[H_g,\hat \rho]+R_{\mathrm{se}}(\hat \varphi(\mathbb{I}+4\langle\mathbf S\rangle\cdot\mathbf S)-\hat{\rho})\\
&+R_{\mathrm{wall}}\left(\mathbb{I}-\hat{\rho}\right)+R_{\mathrm{sd}}(\hat{\varphi}-\hat \rho)\\
&+R_{\mathrm{op}}(\hat{\varphi}(\mathbb{I}+2\mathbf{s}\cdot\mathbf{S})-\hat{\rho})+D\nabla ^{2}\hat{\rho},
\label{eq2}
\end{aligned}
\end{split}     
\end{equation}
where the density matrix $\hat \rho$ is expressed in the $\ket{F,m_F}$ basis, since the Zeeman splitting is much smaller than the hyperfine splitting. To clarify the action of the relaxation terms, we decompose $\hat \rho$ into a nuclear part $\hat \varphi$ and an electronic part $\boldsymbol{\Theta}\cdot\mathbf{S}$ \cite{appelt1998theory}, with
\begin{equation}
\hat \varphi=\frac{1}{4} \hat \rho+\mathbf{S}\hat \varrho\,\mathbf{S}.
\end{equation}
Here $\mathbf{s}=i\boldsymbol{\epsilon}\times\boldsymbol{\epsilon}^{*}$ denotes the average photon spin, where $\boldsymbol{\epsilon}=[\epsilon_x,\epsilon_y,0]^{\mathrm T}$ is the light polarization vector. 

Equation~(\ref{eq2}) contains several processes that degrade the battery capacity.
(1) Spin-exchange collisions (rate $R_{\mathrm{se}}$): these drive the electronic part $\boldsymbol{\Theta}\cdot S$ towards the instantaneous average spin average spin $\langle\mathbf S\rangle$, while leaving the nuclear part $\hat\varphi$ unchanged. As a result, the incoherent capacity remains unaffected, whereas the coherent capacity is reduced. (2) Wall collisions (rate $R_{\mathrm{wall}}$): atoms colliding with the cell wall experience strong, random electromagnetic fields from ions and molecules in the glass, leading to a complete randomization of $\hat{\varrho}$ and driving the battery towards a maximally mixed state. The rate $R_{\mathrm{wall}}$ is related to the diffusion coefficient $D$. In the experiment, this relaxation channel is strongly suppressed by the paraffin coating on the cell walls. (3) Spin-destructive collisions (rate $R_{\mathrm{se}}$): these redistribute the electronic population and drive the system towards a non-polarized state with vanishing capacity. In this process, angular momentum is transferred from the atomic spin to the collective rotational degrees of freedom of the atoms and then be dissipated. (4) Optical pumping (rate $R_{\mathrm{op}}$): this drives the atoms towards a polarized state through the transfer of energy in the form of angular momentum from the light field.

\bigskip
\noindent\textit{Experimental protocols}\\
Figure~\ref{pulsesequence1} summarizes the pulse sequences for three experimental protocols. Each protocol consists of four stages: state initialization, optional coherence control, unitary exploration, and readout. In Protocol 1, hierarchical traversal is applied to a coherent quantum battery: after initialization by $R_y$ and $R_z$ rotations, the capacity is determined by sweeping $R_z$ and $R_x$ pulses to locate energy extrema. Protocol 2 uses the same initial preparation but evaluates the capacity tomographically: the coherent contribution is measured directly, while a gradient-field pulse followed by an $R_x$ rotation isolates the incoherent contribution, enabling full reconstruction. Protocol 3 applies hierarchical traversal to an incoherent (classical) battery and differs from Protocol 1 only in that a GF pulse is introduced during state preparation to erase coherence while preserving the level populations.

\section*{Acknowledgements}
W.Z. was supported by the National Natural Science Foundation of China under Grant No. 12574532 and No. 92476204. H.W. was supported by Zhejiang Provincial Natural Science Foundation of China under Grant No. LY24A050005. Y.X. was supported by the National Natural Science Foundation of China under Grant No.12204386. M.X. was supported by Interdisciplinary Research of Southwest Jiaotong University China Interdisciplinary Research of Southwest Jiaotong University China under Grant No.2682022KJ004, and Innovation Program for Quantum Science and Technology under Grant No. 2021ZD0301601.

\end{document}